\documentclass{article}
\usepackage{amsmath, amssymb, amsfonts}
\title{Comment on ''Accelerated observers emerging from a Bose-Einstein condensate through analogue gravity''}  
\author{Hristu Culetu, \\Ovidius University, Dept.of Physics, \\ Mamaia Avenue 124, 900527 Constanta, Romania, \\e-mail : hculetu@yahoo.com}

\begin{document}
\numberwithin{equation}{section}
\pagenumbering{arabic}
\maketitle
\newcommand{\fv}{\boldsymbol{f}}
\newcommand{\tv}{\boldsymbol{t}}
\newcommand{\gv}{\boldsymbol{g}}
\newcommand{\OV}{\boldsymbol{O}}
\newcommand{\wv}{\boldsymbol{w}}
\newcommand{\WV}{\boldsymbol{W}}
\newcommand{\NV}{\boldsymbol{N}}
\newcommand{\hv}{\boldsymbol{h}}
\newcommand{\yv}{\boldsymbol{y}}
\newcommand{\RE}{\textrm{Re}}
\newcommand{\IM}{\textrm{Im}}
\newcommand{\rot}{\textrm{rot}}
\newcommand{\dv}{\boldsymbol{d}}
\newcommand{\grad}{\textrm{grad}}
\newcommand{\Tr}{\textrm{Tr}}
\newcommand{\ua}{\uparrow}
\newcommand{\da}{\downarrow}
\newcommand{\ct}{\textrm{const}}
\newcommand{\xv}{\boldsymbol{x}}
\newcommand{\mv}{\boldsymbol{m}}
\newcommand{\rv}{\boldsymbol{r}}
\newcommand{\kv}{\boldsymbol{k}}
\newcommand{\VE}{\boldsymbol{V}}
\newcommand{\sv}{\boldsymbol{s}}
\newcommand{\RV}{\boldsymbol{R}}
\newcommand{\pv}{\boldsymbol{p}}
\newcommand{\PV}{\boldsymbol{P}}
\newcommand{\EV}{\boldsymbol{E}}
\newcommand{\DV}{\boldsymbol{D}}
\newcommand{\BV}{\boldsymbol{B}}
\newcommand{\HV}{\boldsymbol{H}}
\newcommand{\MV}{\boldsymbol{M}}
\newcommand{\be}{\begin{equation}}
\newcommand{\ee}{\end{equation}}
\newcommand{\ba}{\begin{eqnarray}}
\newcommand{\ea}{\end{eqnarray}}
\newcommand{\bq}{\begin{eqnarray*}}
\newcommand{\eq}{\end{eqnarray*}}
\newcommand{\pa}{\partial}
\newcommand{\f}{\frac}
\newcommand{\FV}{\boldsymbol{F}}
\newcommand{\ve}{\boldsymbol{v}}
\newcommand{\AV}{\boldsymbol{A}}
\newcommand{\jv}{\boldsymbol{j}}
\newcommand{\LV}{\boldsymbol{L}}
\newcommand{\SV}{\boldsymbol{S}}
\newcommand{\av}{\boldsymbol{a}}
\newcommand{\qv}{\boldsymbol{q}}
\newcommand{\QV}{\boldsymbol{Q}}
\newcommand{\ev}{\boldsymbol{e}}
\newcommand{\uv}{\boldsymbol{u}}
\newcommand{\KV}{\boldsymbol{K}}
\newcommand{\ro}{\boldsymbol{\rho}}
\newcommand{\si}{\boldsymbol{\sigma}}
\newcommand{\thv}{\boldsymbol{\theta}}
\newcommand{\bv}{\boldsymbol{b}}
\newcommand{\JV}{\boldsymbol{J}}
\newcommand{\nv}{\boldsymbol{n}}
\newcommand{\lv}{\boldsymbol{l}}
\newcommand{\om}{\boldsymbol{\omega}}
\newcommand{\Om}{\boldsymbol{\Omega}}
\newcommand{\Piv}{\boldsymbol{\Pi}}
\newcommand{\UV}{\boldsymbol{U}}
\newcommand{\iv}{\boldsymbol{i}}
\newcommand{\nuv}{\boldsymbol{\nu}}
\newcommand{\muv}{\boldsymbol{\mu}}
\newcommand{\lm}{\boldsymbol{\lambda}}
\newcommand{\Lm}{\boldsymbol{\Lambda}}
\newcommand{\opsi}{\overline{\psi}}
\renewcommand{\tan}{\textrm{tg}}
\renewcommand{\cot}{\textrm{ctg}}
\renewcommand{\sinh}{\textrm{sh}}
\renewcommand{\cosh}{\textrm{ch}}
\renewcommand{\tanh}{\textrm{th}}
\renewcommand{\coth}{\textrm{cth}}

\begin{abstract}
Few comments upon Gonzalez-Fernandez and Camacho paper (arXiv: 1904.02299) are pointed out. We bring evidences that the analogue gravity recipe does not work for a BEC in an anisotropic harmonic-oscillator trap. The analogy with an accelerated observer does not seem to be realistic. Some loopholes related to the physical units are emphasized.
 \end{abstract}

  Most textbooks on Statistical Mechanics treat the phenomenon of Bose-Einstein condensate (BEC) in an uniform, non-interacting gas of bosons. In the semi-classical approximation, the energy spectrum is considered as a continuum. In the fully-condensed state, all bosons are in the same single-particle state \cite{PS}.
	
	Gonzalez-Fernandez and Camacho \cite{GFC} use the Analogue Gravity recipe in a study on moving fluids and an analog spacetime based on the acoustics of the fluid. The BEC is trapped in an anisotropic 3-dimensional harmonic-oscillator potential. In the acoustic representation, the authors' effective metric $g_{\mu \nu}(t,\vec{r})$ (their Eq.4) contains the prefactor $n_{c}/mc_{s}$, where $c_{s}$ is the phonons speed in the medium, $m$ is the mass of the particles of the BEC and $n_{c}$ is the density distribution. 
	
	From authors' Gross-Pitaevskii equation (1) (or from their Eq.2) we notice that the coupling constant $k(a)$, characterising the effective interaction between the particles, has units of energy $\times$ volume while $n_{c}$ is 1/(volume). Therefore, the prefactor $n_{c}/mc_{s}$ from their Eq.4 has the dimension $T M^{-1} L^{-4}$ ( T- time, M - mass, L - length) (see also Eq.13). But $g_{\mu \nu}$ - the metric tensor in General Relativity (GR) should be dimensionless or a length squared if we take the coordinates to be dimensionless. In either case, it turns out that the units are erronous, even though the authors use geometrical units ($C = 1$, where $C$ is the velocity of light in vacuo). Having dimensions, we cannot absorb the prefactor in the velocities $c_{s}$ and $v$.\footnote{At p.6, the authors introduced a $g_{\mu \nu}$ with the right dimensions (Eq.16) by simply getting rid of a (non dimensionless) conformal factor.} In addition, the spacetime (4) has no an event horizon because there is no any value of $v^{2} = v_{i}v^{i}$ to satisfy the equation \cite{CGLP, HC}
\begin{equation}
g_{tt} - \sum \frac{g_{ti}^{2}}{g_{ii}} = 0,~~~i = 1, 2, 3
 \label{1}
 \end{equation}
	which is the condition to have a horizon (the l.h.s. of (0.1) equals $-c_{s}^{2}$, that is nonzero). 
	
	Let us observe that $K$ from Eq.17 has the correct dimensions - a velocity squared. However, when $x <<b, y<<c$ and $z<<d$, the exponential factor tends to unity and $K = c_{s}^{2}$, using (10). Therefore, the metric is not Minkowskian since the constant $c_{s}$ is not Lorentz-invariant, like the speed of light ($ds^{2}$ will not be invariant under the Lorentz transformation). In other words, we may not define the fourth coordinate as $x^{0} = c_{s}t$. In our view, there is no way to get the Minkowski geometry from the metric (17) because $K$ cannot become the velocity of light in any approximation, even in the case the BEC is removed. 
	
	We express serious doubts on the authors' physical interpretation of the line-element (17). The metric (18) is nothing but the Rindler metric which is flat but covers only a part of Minkowski's spacetime. As the authors use geometrical units - see below Eq.27 - their $g_{00}$ in (18) is, in fact, $g_{00} = -1-g\xi^{1}/C^{2}$. Same is valid for Eq. (20) and (21). $g$ is the constant acceleration of an observer located at the origin $\xi^{1} = 0$ of the accelerated reference system or the surface gravity
\begin{equation}
\kappa = \sqrt{a^{b}a_{b}} \sqrt{-g_{00}}|_{\xi^{1} = -1/g} = \frac{g}{1+g\xi^{1}} (1+g\xi^{1}) = g,
 \label{2}
 \end{equation}
with $a^{b} = (0, g/(1+g\xi^{1}), 0, 0)$ the acceleration of a static observer and $\xi^{1} = -1/g$ is the location of the Rindler horizon.

As far as the generalized form (21) of the approximated line-element (20) is concerned, it is worth noting that the spatial components of the acceleration are not $g_{i}$, as the authors of \cite{GFC}	claim, but $g_{i}/(1+2g_{i}\xi^{i})$. Moreover, we stress that the line-elements (20) and (21) are curved, though the starting Rindler metric (18) was flat. In other words, the authors generated a stress tensor by means of an approximation ($g\xi^{1}/C^{2} <<1)$, which is not reasonable from the physical point of view. We also notice that $t'$ from $dt' = \sqrt{K} dt$ is not a time, but a length ($\sqrt{K}$ is a velocity). 
The points inside the harmonic-oscillator trap obey, of course, $(x_{i}/b_{i})<1$ but from here we cannot conclude that $(x_{i}^{2}/b_{i}^{2})<<1$, as the authors have claimed in Eq.23 ( from, say, $x_{i}/b_{i} = 0.9 <1$ one may not deduce that $x_{i}^{2}/b_{i}^{2} = 0.81 <<1$). 

As another evidence that the implication (23) is not always valid, one observes that the spatial accelerations of a static observer in the spacetime (21) are $g_{i}/(1+2g_{i}\xi^{i})$, which have a very different behavior compared to $-x_{i}/2b_{i}^{2}$. How do the authors ensure the same physical units in their Eq.25 ? The l.h.s. is an acceleration but the r.h.s has dimension $1/$length. Because they used geometrical units, the l.h.s. should be $g_{i}/C^{2}$. But there is no any $C^{2}$ in the metric coefficients of (24), since $dt'^{2} = K dt^{2}$. It turns out there is a contradiction here.

  Our conclusion is therefore that the analogue-gravity system to that of a BEC in an anisotropic harmonic-oscillator trap cannot be considered as an accelerated observer.
	
	Concerning the Sec.3, the stress tensor (32) for a spherically-symmetric trap does not seem to be realistic. For instance, taking $r<<b$ in (32) (a case of a more interest), we find that $T_{\mu \nu}^{s}= (C^{4}/8\pi G) G_{\mu \nu}^{s}$ is of the order of $C^{4}/8\pi Gb^{2}$, where the width of the wave function $b = \sqrt{\hbar/m\omega}$ (see \cite{PS}, Eq.2.34), with $\omega$ - the trap frequency. If we consider $m = 1 amu$ and $\omega = 10^{5}$ Hz, one obtains $b = 1 \mu m$, whence a component of the energy-momentum tensor, say the energy density, gives $10^{44} ergs/(\mu m)^{3}$, an unphysical value. Similar conclusions can be drawn for the axially-symmetric trap and for the asymmetric trap.
	
	To summarize, although the paper is interesting and innovative, we have brought evidences that the analogue gravity applied to a Bose-Einstein condensate in a harmonic-oscillator trap would not work and the system may not be regarded as an accelerated observer.


\begin{thebibliography} {2}

\bibitem{PS}
C. J.Pethick and H. Smith, \textit{Bose-Einstein condensate in diluted gases}, Cambridge University Press, 2002.
\bibitem{GFC}
B. Gonzalez-Fernandez and A. Camacho, arXiv: 1904.02299.
\bibitem{CGLP}
M. Cvetic, G. W. Gibbons, H. Lu and C. N. Pope, arXiv: hep-th/0504080.
\bibitem{HC}
H. Culetu, J. Phys. Conf. Ser.68 (2007) 012036; arXiv: hep-th/0602014.




\end{thebibliography}
\end{document}